\documentclass[prd,nofootinbib,showpacs]{revtex4}
\usepackage{amsmath}
\usepackage{graphicx}
\usepackage[dvipdfm,colorlinks]{hyperref}
\usepackage{color} 
\usepackage{epsfig}
\usepackage{epsf}

\begin{document}
\title{Astrophysical Constraints on the scale of Left-Right Symmetry in Inverse Seesaw Models \\}
\author{Debasish Borah}
\email{debasish@tezu.ernet.in}
\affiliation{Department of Physics, Tezpur University, Tezpur - 784028, India}

\begin{abstract}
We revisit the recently studied supersymmetric gauged inverse seesaw model \cite{An:2011uq} to incorporate astrophysical constraints on lightest supersymmetric particle (LSP) lifetime such that LSP constitutes the dark matter of the Universe. The authors in \cite{An:2011uq} considered light sneutrino LSP that can play the role of inelastic dark matter (iDM) such that desired iDM mass splitting and tiny Majorana masses of neutrinos can have a common origin. Here we consider a generalized version of this model without any additional discrete symmetry. We point out that due to spontaneous R-parity $(R_p = (-1)^{3(B-L)+2s})$ breaking in such generic supersymmetric gauged inverse seesaw models, LSP can not be perfectly stable but decays to standard model particles after non-renormalizable operators allowed by the gauge symmetry are introduced. We show that strong astrophysical constraints on LSP lifetime makes sneutrino dark matter more natural than standard neutralino dark matter. We also show that long-livedness of sneutrino dark matter constrains the left right symmetry breaking scale $M_R < 10^4 \; \text{GeV}$. 
\end{abstract}

\maketitle

\section{Introduction}
Left-Right Symmetric Models (LRSM) \cite{Pati:1974yy,Mohapatra:1974gc, Senjanovic:1975rk, 
Mohapatra:1980qe, Deshpande:1990ip} provide a framework within which spontaneous parity breaking as well as tiny neutrino masses \cite{Fukuda:2001nk, Ahmad:2002jz, 
Ahmad:2002ka, Bahcall:2004mz} can be successfully implemented without reference to very high scale physics such as grand unification. Incorporating
Supersymmetry (SUSY) into it comes with other advantages like
providing a solution to the gauge hierarchy problem, and providing 
a Cold Dark Matter candidate which is the lightest
supersymmetric particle (LSP). In Minimal Supersymmetric
Standard Model (MSSM), the stability of LSP is guaranteed by R-parity,
defined as $R_p =(-1)^{3(B-L)+2S}$ where $S$ is the spin of the particle.
This is a discrete symmetry put by hand in MSSM to keep the baryon
number (B) and lepton number (L) violating terms away from the
superpotential. In generic implementations of Left-Right
symmetry, R-parity is a part of the gauge symmetry and hence not ad-hoc like in
the MSSM. In one class of models ~\cite{Aulakh:1998nn, Aulakh:1997ba,Babu:2008ep,Patra:2009wc}, spontaneous parity breaking is 
achieved without breaking R-parity. This was not possible in minimal supersymmetric left right (SUSYLR) models where the only way to break parity is to consider 
spontaneous R-parity violation \cite{Kuchimanchi:1993jg}. In minimal
SUSYLR model parity, $SU(2)_R$ gauge symmetry as well as R-parity break 
simultaneously by the vacuum expectation value of right handed sneutrino. 

Here we study a different SUSYLR model which belong to a more general 
class of models where both R-parity and D-parity break spontaneously \cite{Borah:2011zz} by the vacuum expectation value (vev) of a Higgs field carying $U(1)_{B-L}$ gauge charge $\pm 1$ and hence odd under R-parity. Spontaneous R-parity breaking models have received lots of attention recently due to their rich phenomenology \cite{FileviezPerez:2008sx,Barger:2008wn,FileviezPerez:2012mj}. In such generic spontaneous R-parity breaking models, the scalar superpartner of right-handed neutrino acquire a non-zero vev which breaks $U(1)_{B-L}$ symmetry spontaneously. Such a scenario gives rise to tree level mixing between neutralinos and light neutrinos and hence the neutralino dark matter candidate is lost in such a model unless one talks about long lived gravitino dark matter. However the model we study in this letter, although breaks R-parity spontaneously, does not give rise to tree level mixing terms between LSP and standard model fermions. Thus we can have a dark matter candidate in such a model without introducing the least understood gravity sector into account. Recently right handed sneutrino dark matter in such a model was discussed in \cite{An:2011uq}. However, the authors in \cite{An:2011uq} (also in \cite{BhupalDev:2010he}) considered an additional discrete $Z_2$ symmetry so as to guarantee a perfectly stable LSP. Here we consider a generalized version of this model without any additional symmetries apart from the gauge symmetry. We point out that LSP dark matter, although stable at the renormalizable level, decays after higher dimensional gauge invariant terms are introduced. The strength of such operators will be tightly constrained from the fact that LSP lifetime should be longer than the age of the Universe and large enough so as to agree with astrophysical observations of nearby galaxies and clusters \cite{Dugger:2010ys}. Astrophysical constraints on such operators within the framework of MSSM was studied in \cite{Berezinsky:1996pb}. Here we follow a similar analysis in our model and show that astrophysical constraints not only put an upper bound on the left-right symmetry breaking scale but also make the sneutrino dark matter more natural than standard neutralino dark matter. It is worth mentioning that constraints on the left-right symmetry breaking scale in such a model were derived recently in \cite{Borah:2011qq} from the requirement of successful gauge coupling unification and disappearance of transitory domain walls formed as a result of spontaneous discrete symmetry breaking.

This letter is organized as follows. In section \ref{sec:SUSYLR} we briefly review the model. In section \ref{sec:High} we discuss the higher dimensional operators in the model and astrophysical constraints. We summarise the constraints from gauge coupling unification and domain wall disappearance from our earlier work \cite{Borah:2011qq} in section \ref{sec:dwunif} and finally conclude in section \ref{sec:con}.

\section{The Model}
\label{sec:SUSYLR}
Spontaneous R-parity breaking can be achieved even without 
giving vev to the sneutrino fields. If the $U(1)_{B-L}$ symmetry is broken
by a Higgs field which has odd $B-L$ charge then R-parity is spontaneously broken. We call this model as Minimal Higgs Doublet (MHD) Model. The minimal such model \cite{Borah:2009ra,Borah:2011zz} has the following particle content
\begin{equation}
L(2,1,-1), \quad L_c(1,2,1), \quad S (1,1,0), \quad
Q(2,1,\frac{1}{3}),\quad  Q_c(1,2, -\frac{1}{3}) \nonumber
\end {equation}
\begin{equation}
H=
\left(\begin{array}{cc}
\ H^+_{L} \\
\ H^0_{L}/{\surd 2}
\end{array}\right)
\sim (2,1,1), \quad
H_c=
\left(\begin{array}{cc}
\ H^+_{R} \\
\ H^0_{R}/{\surd 2}
\end{array}\right)
\sim (1,2,-1), \nonumber
\end{equation}
\begin{equation}
\bar{H}=
\left(\begin{array}{cc}
\ h^0_{L}/{\surd 2} \\
\ h^-_{L}
\end{array}\right)
\sim (2,1,-1), \quad
\bar{H}_c=
\left(\begin{array}{cc}
\ h^0_{R}/{\surd 2} \\
\ H^-_{R}
\end{array}\right)
\sim (1,2,1), \nonumber
\end{equation}
\begin{equation}
\Phi_1(2,2,0), \quad \Phi_2(2,2,0) \nonumber 
\end{equation}
where the numbers in brackets correspond to the quantum numbers corresponding 
to $ SU(2)_L\times SU(2)_R \times U(1)_{B-L} $. The symmetry breaking pattern 
is 
\begin{eqnarray}
SU(2)_L \times SU(2)_R \times U(1)_{B-L} \quad
\underrightarrow{\langle H,H_c \rangle} \quad SU(2)_L \times U(1)_{Y} \quad
\underrightarrow{\langle \Phi \rangle}\quad U(1)_{em}
\end{eqnarray}

Neutrino masses arise naturally in this model by so called inverse seesaw mechanism by virtue of the presence of singlet superfields $S(1,1,0)$ (one per generation). The renormalizable superpotential relevant for the spontaneous parity violation and neutrino mass is given as follows
\begin{eqnarray}
%\lefteqn{
\lefteqn{W_{ren}=h^{(i)}_l L^T\tau_2 \Phi_i \tau_2 L_c+ h^{(i)}_q Q^T\tau_2 
\Phi_i \tau_2 Q_c + \iota fL^T \tau_2 S H +\iota f^*L^T_c 
\tau_2 S H_c+ M_s SS} \nonumber \\
&& +\mu_{ij}\text{Tr}\tau_2\Phi^T_i\tau_2\Phi_j + 
f_{h\phi}(H^T\Phi_i H_c+\bar{H}^T\Phi_i \bar{H}_c)+ m_h H^T 
\tau_2 \bar{H} +m_h H^T_c \tau_2 \bar{H}_c
\end{eqnarray}

We denote the vev of the neutral components of $\Phi_1, 
\Phi_2, H_L, \bar{H}_L, H_R, \bar{H}_R$ as $\langle (\Phi_1)_{11} 
\rangle = v_1,~ \langle (\Phi_2)_{22} \rangle = v_2,~ \langle H_L,\bar{H}_L \rangle 
= v_L,~ \langle H_R,\bar{H}_R \rangle =v_R$

The neutrino mass matrix in the basis $(\nu, \nu_c, S)$ is given by
\begin{equation}
M_{\nu \rho} =
\left(\begin{array}{cccc}
\ 0 & M_D &  F v_L \\
\ M^T_D & 0 & F' v_R\\
\ F^T v_L & F'^T v_R  & M_s
\end{array}\right)
\end{equation}
where $M_D = (\phi^0_{12}h_1+\phi^0_{22}h_2), F = f/\sqrt{2}, F' = f^*/\sqrt{2} $. After orthogonalization we get the following expression for $ \nu $ mass 
\begin{equation}
M_{\nu} = - M_D M_R^{-1} M_D^T - \left( M_D + M_D^T \right)v_{L}/v_{R} 
\label{nu1}
\end{equation}
where 
\begin{equation}
M_R = (F' \,v_{R}) {M_s}^{-1} (F'^T \,v_{R})
\label{nu2}
\end{equation}
It should be noted from the neutrino mass matrix that these mass terms allow the mixing of an R-parity odd singlet fermion $S$ with an R-parity even neutrino. Note that the superpotential preserves R parity. The mild R parity
violation occurring in the
neutrino mass matrix should be understood as an accidental consequence of
$B-L$ gauge symmetry
breakdown. 

Neutrino mass can arise from type III seesaw mechanism \cite{Foot:1988aq} if we introduce fermionic triplets instead of singlets. However when we have a TeV scale intermediate $U(1)_{B-L}$ symmetry, the fermion triplets will spoil the gauge coupling unification \cite{Borah:2010zq,Borah:2011zz} and hence fermion singlets will serve a better purpose in this case.

\section{Non-renormalizable operators and Astrophysical Constraints}
\label{sec:High}

The authors of \cite{Berezinsky:1996pb} considered explicit R-parity violating terms in the MSSM superpotential of the form $\lambda (L L E^c) + \epsilon (LH)$ which lead to the decay of LSP dark matter candidate in the model. 
Similar analysis within the framework of grand unified theories can be found in \cite{Arvanitaki:2008hq,Arvanitaki:2009yb}. We take the conservative lower bound $(\tau_{LSP} > 10^{27} \; \text{s})$ on LSP lifetime coming from the recent Fermi telescope observation of nearby galaxy and clusters \cite{Dugger:2010ys}.

In the model we are studying, the effective terms in the superpotential leading to LSP decay can arise after introduing dimension four and dimension five operators as follows:

$$ W_{non-ren} \supset \frac{f_1}{\Lambda} (L^T \tau_2 H + L^T_c \tau_2 H_c) (H^T \tau_2 \bar{H} + H^T_c \tau_2 \bar{H}_c) + \frac{f_2}{\Lambda^2} (L^T\tau_2 \Phi_i \tau_2 L_c)(L^T \tau_2 H + L^T_c \tau_2 H_c) $$

The first term give rise to terms like $\epsilon (L H)$ in the low energy effective theory after gauge symmetry is spontaneously broken. The strength of such a term is dictated by $ \epsilon_{non-ren} \sim \langle H_c \rangle^2/\Lambda $. Here $\langle H_c \rangle $ is the left-right symmetry breaking scale which has a lower bound of $M_{W_R} \geq 2.5 \; \text{TeV}$ \cite{Maiezza:2010ic}. And, the cut-off scale $\Lambda$ is the generic grand unified theory (GUT) scale $\Lambda = \Lambda_{GUT} \sim 2 \times 10^{16} \; \text{GeV}$. Using these values and assuming generic order one dimensionless coefficients $f_1$, we have
\begin{equation}
\epsilon_{non-ren}  > 10^{-10} \; \text{GeV}
\label{nonren1}
\end{equation}

As shown in \cite{Berezinsky:1996pb}, the decay width of neutralino corresponding to a term $\epsilon (LH)$ in the superpotential is given by
\begin{equation}
\Gamma_{\chi} \propto \epsilon^2 \frac{G^2_F m^3_{\chi}}{768 \pi^3}
\end{equation}
with constant of proportionality of order unity. Now, for generic neutralino dark matter with mass of the order of $100 \; \text{GeV}$, the astrophysical constraint on LSP lifetime $(\tau_{LSP} > 10^{27} \; \text{s})$ gives rise to
\begin{equation}
\epsilon_{astro} < 10^{-22} \; \text{GeV}
\label{astro1}
\end{equation}

Clearly the astrophysical bound (\ref{astro1}) does not agree with the strength of $\epsilon_{non-ren}$ arising from generic non-renormalizable operators in the theory. If we fine tune $f_1$ to be as small as electron Yukawa coupling $10^{-5}$, then $\epsilon$ can be as small as $10^{-15}$. But this lies around seven orders of magnitude above the upper bound set by astrophysical constraints (\ref{astro1}). Thus, standard neutraino dark matter is very unlikely in these models unless we have unnatural fine tuning of the dimensionless coefficients in the non-renormalizable operators. It should be noticed that a term like $\epsilon (L H)$ arise at tree level in generic spontaneous R-parity violating models with non-zero right handed sneutrino vev \cite{FileviezPerez:2008sx,Barger:2008wn,FileviezPerez:2012mj}.

The second term in the non-renormalizable superpotential gives rise to an effective term of the form $\lambda_{non-ren} (L L_c L_c) $ which opens the decay channel of sneutrino into two standard model fermions. The strength of such a term is given by $\lambda_{non-ren} \sim \langle \phi \rangle \langle H_c \rangle /\Lambda^2 $ where $\langle \phi \rangle \sim 10^2 \; \text{GeV}$ and $\langle H_c \rangle > 2.5 \; \text{TeV}$. For $\Lambda = \Lambda_{GUT}$ such a term is of strength
\begin{equation}
\lambda_{non-ren} > 10^{-27} 
\label{nonren2}
\end{equation}
The decay width of a sneutrino to standard model fermion-antifermion pairs is given by 
\begin{equation}
\Gamma_{\tilde{\nu}} = \frac{\lambda^2 m_{\tilde{\nu}}}{8\pi} \left (1-\frac{4m^2_f}{m^2_{\tilde{\nu}}} \right )^{3/2}
\end{equation}
Now, for sneutrino LSP mass of the order of $100 \; \text{GeV}$, the astrophysical constraint on LSP lifetime $(\tau_{LSP} > 10^{27} \; \text{s})$ gives rise to
\begin{equation}
\lambda_{astro} < 10^{-26}
\label{astro2}
\end{equation}
which agrees with the generic $\lambda_{non-ren}$ arising from the non-renormalizable operators in the theory (\ref{nonren2}). Thus sneutrino LSP in such a model can be a viable dark matter candidate provided it satisfies other relevant constraints of relic density, direct detection etc. Recently it was shown that such a right handed sneutrino dark matter (within the framework of a similar left right model) can satisfy relic density as well as direct detection constraints \cite{An:2011uq}.

For right handed sneutrino dark matter to obey the relevant astrophysical constraints (\ref{astro2}), the left right symmetry breaking scale should however have an upper bound. Requiring $ \langle \phi \rangle \langle H_c \rangle /\Lambda_{GUT}^2 < 10^{-26} $ gives rise to a bound on the left-right symmetry breaking scale 
\begin{equation}
\langle H_c \rangle < 10^4 \; \text{GeV}
\end{equation}
for generic GUT scale and order one dimensionless couplings. However, as studied in \cite{Borah:2011zz,Borah:2011qq} and summarised in the next section, successful gauge coupling unification in such a minimal model puts a lower bound on left-right symmetry breaking scale $\sim 10^{12} \; \text{GeV}$.

\begin{figure}
\begin{center}
\includegraphics{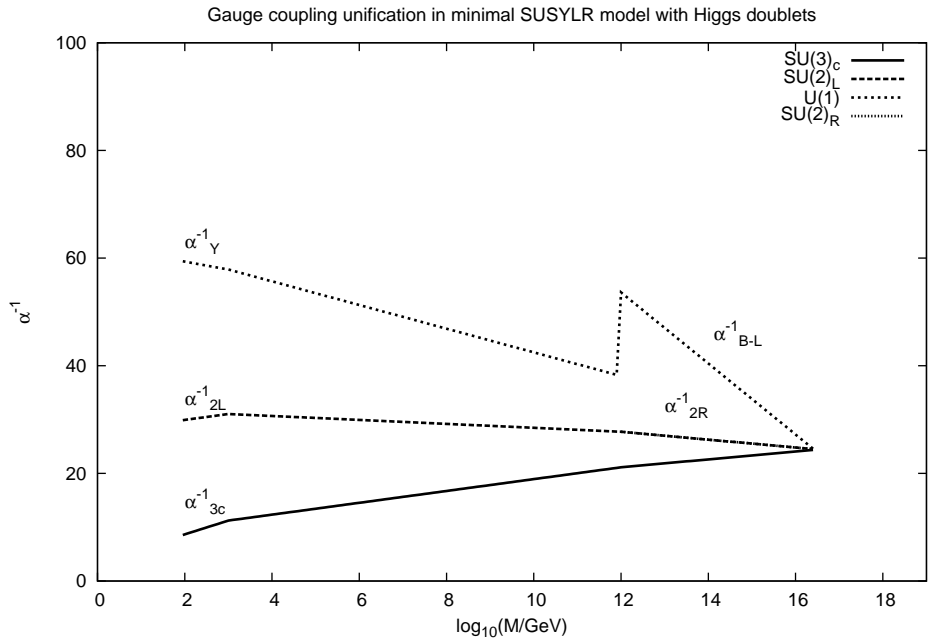}
\end{center}
\caption{Gauge coupling unification in minimal SUSYLR model with Higgs doublets, $M_{susy} = 1$ TeV, 
$M_R = 10^{12}$ GeV, $M_{GUT} = 10^{16.4}$ GeV. The figure is redrawn from \cite{Borah:2011zz,Borah:2011qq}}
\label{fig1}
\end{figure}
\begin{figure}
\begin{center}
\includegraphics{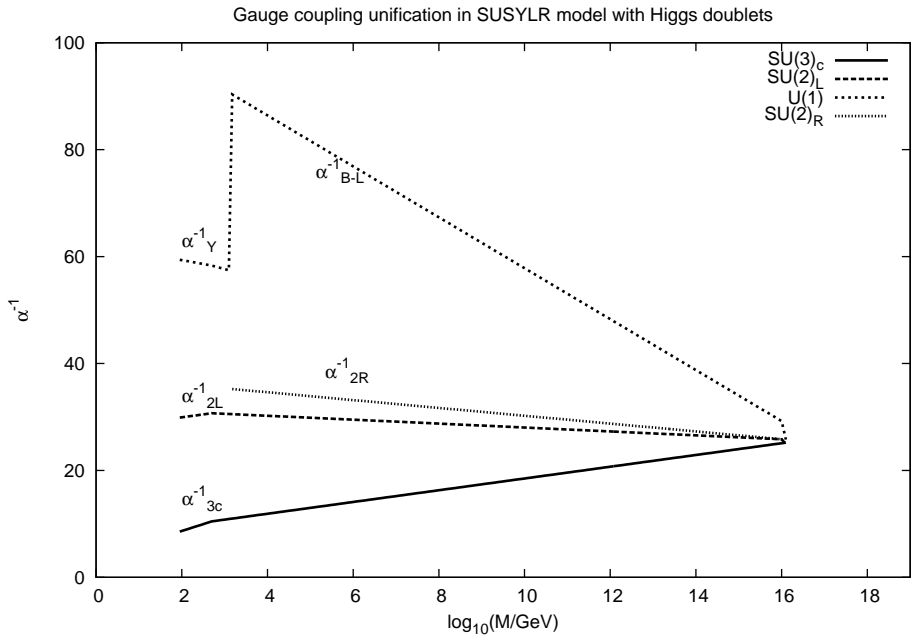}
\end{center}
\caption{TeV scale $M_R$ is possible with the introduction of parity odd singlets into the MHD model. The figure is redrawn from \cite{Borah:2010zq,Borah:2011zz} with $M_{susy} = 500 \; \text{GeV}$ and $M_{GUT} = 10^{16} \; \text{GeV}$.}
\label{fig2}
\end{figure}

\section{Constraints on $M_R$ from Unification and Domain Wall Disappearance}
\label{sec:dwunif}
Similar to generic SUSYLR models, here also 
the intermediate symmetry breaking scales are constrained by 
demanding successful gauge coupling unification at a very high
 scale $M_G ( >2 \times 10^{16} \text{GeV})$. The couplings of $U(1)_{B-L}$ and $SU(2)_{L,R}$ meet much 
before the allowed Unification scale if the intermediate symmetry 
breaking scale $M_R$ is lower than a certain value. For the minimal
 SUSYLR model with Higgs doublets, this lower bound on $M_R$ is found 
to be of the order of $10^{12}$ GeV. We also consider two additional 
heavy colored superfields so that the $SU(3)_c$ coupling meet the other 
two couplings at one point. They are denoted as $\chi(3,1,1,-\frac{2}{3}), 
\bar{\chi}(\bar{3},1,1,\frac{2}{3})$ and can be accommodated within $SO(10)$ 
GUT theory in the representations $\textbf{120},\overline{\textbf{126}}$. Here 
we assume that the structure of the GUT theory is such that these fields 
survive the symmetry breaking and can be as light as the $SU(2)_R$ breaking scale. 
The resulting gauge coupling unification as shown in the figure \ref{fig1}.

As discussed in details in \cite{Borah:2011qq}, the succesful disappearance of domain walls
in this model do not put any strict constraints on the left-right symmetry breaking scale and can be anywhere
between a TeV scale and the Planck scale. Thus unification and domain wall disappearance constraints are 
compatible with each other. The discrepancy between the astrophysical limit $M_R < 10^4 \; \text{GeV}$ and the limit from successful unification $M_R \geq 10^{12} \text{GeV}$ can be removed by including a parity odd singlet to our model. As studied in \cite{Dev:2009aw,Borah:2010zq,Borah:2011zz}, such a framework allows even a TeV scale $M_R$ from the requirement of successful gauge coupling unification as can be seen from figure \ref{fig2}. It should be noted that the authors of \cite{An:2011uq} indeed considered such a model with parity odd singlet which allows a TeV scale $M_R$. Such TeV scale $M_R$ is not just a requirement from astrophysical constraints as we have found above, but these TeV scale gauge bosons also contribute to the dark matter annihilations \cite{An:2011uq} in the early universe producing the correct relic density at present.

\section{Results and Conclusion}
\label{sec:con}
We have discussed the issue of stability of LSP dark matter in a specific version of SUSYLR model with inverse seesaw mechanism of neutrino mass where both D-parity and R-parity are spontaneously broken. We point out that, although LSP is a stable particle 
in the renormalizable version of the model, it can decay into standard model fermions after the non-renormalizable terms are introduced. The requirement that LSP dark matter should be long lived so as to satisfy strict astrophysical and cosmological bounds constrains the strength of these higher dimensional operators suppressed by GUT scale. We point out that standard neutralino dark matter (decaying through dimension four operators in the superpotential) scenario is disfavored in this model unless one considers unnatural fine-tuning of the dimensionless coefficients in the higher dimensional operators. However, right handed sneutrino dark matter (decaying through dimension five operators in the superpotential) satisfy the astrophysical bounds more naturally and can be a viable dark matter candidate provided it satisfies other relevant constraints like relic density, direct detection etc.

Interestingly, the dimension five operators leading to sneutrino decay involve the left right symmetry breaking scale. The requirement that the strength of such an operator should be small enough to satisfy astrophysical bounds constrains the left right symmetry breaking scale. For generic GUT scale and order one dimensionless couplings, we find this bound to be $M_R < 10^4 \; \text{GeV}$. However, as studied in \cite{Borah:2011zz,Borah:2011qq}, successful gauge coupling unification puts a lower bound $M_R \geq 10^{12} \; \text{GeV}$. The mismatch between these two bounds can be fixed by introducing a parity odd singlet \cite{Dev:2009aw,Borah:2010zq,Borah:2011zz} which allow $M_R$ to be as low as a TeV from the requirement of successful gauge coupling unification. Such TeV scale gauge bosons, apart from satisfying the astrophysical constraints also opens up new dark matter annihilation channels \cite{An:2011uq} producing the correct relic density in the present Universe.

\section{Acknowledgement}
I would like to thank Prof Urjit A. Yajnik, IIT Bombay for useful comments and discussions.

%
%  % annh_rho.eps: 0x0 pixel, 300dpi, 0.00x0.00 cm, bb=154 567 349 693

%\bibliographystyle{apsrev}
%\bibliography{refdparity_dwall_rpv,ref_dw_rpv,spv_dm}
%\bibliography{lrsm,susy,lrsusy,gaugebk,higgs,cosmology,genesis,neutrino,sm,med,sugra}

\end{document}